\documentclass[aps,11pt,showpacs,showkeys]{revtex4}
\usepackage[dvips]{graphicx}

\begin{document}
\title{The low-energy electronic structure \\
and the orbital magnetism in NiO$^\spadesuit$}
\author{R. J. Radwanski}
\affiliation{Center of Solid State Physics, S$^{nt}$Filip 5, 31-150 Krakow, Poland,\\
Institute of Physics, Pedagogical University, 30-084 Krakow,
Poland}
\author{Z. Ropka}
\affiliation{Center of Solid State Physics, S$^{nt}$Filip 5,
31-150 Krakow, Poland}
\homepage{http://www.css-physics.edu.pl}
\email{sfradwan@cyf-kr.edu.pl}

\begin{abstract}
The orbital and spin moment of the Ni$^{2+}$ ion in NiO has been
calculated within the quasi-atomic approach. The orbital moment of
0.46 $\mu_{B}$ amounts at 0 K, in the magnetically-ordered state, to
about 20\% of the total moment (2.45 $\mu_{B}$). For this outcome,
being in nice agreement with the recent experimental finding of the
orbital moment, taking into account the intra-atomic spin-orbit
coupling is indispensable.

\end{abstract}
\maketitle\vspace {-0.5cm}

{\bf 1. Introduction}

NiO attracts large attention of the magnetic community by more
than 50 years. Despite of its simplicity (two atoms, NaCl
structure, well-defined antiferromagnetism (AF) with T$_{N}$ of
525 K) and enormous theoretical and experimental works the
consistent description of its properties, reconciling
its insulating state with the unfilled 3$d$ band is still not reached \cite%
{1,2,3,4,5}.

The aim of this paper is to report the calculations of the magnetic moment
of NiO. We attribute this moment to the Ni$^{2+}$ ions. We have calculated
the moment of the Ni$^{2+}$ ion, its spin and orbital parts, in the NiO$_{6}$
octahedral complex and the orbital moment as large as 0.46 $\mu _{B}$ at 0 K
has been revealed. The approach used can be called the quasi-atomic approach
as the starting point for the description of a solid is the consideration of
the atomic-like electronic structure of the constituting atoms/ions, in the
present case of the Ni$^{2+}$ ions.

{\bf 2. Theoretical outline}

We have treated the 8 outer electrons of the Ni$^{2+}$ ion as forming the
highly-correlated electron system 3d$^{8}$. Its ground term is described by
two Hund's rules yielding $S$=1 and $L$=3, i.e. the ground term $^{3}F$ \cite%
{6}. Such the localized highly-correlated electron system
interacts in a solid with the charge and spin surroundings. The
charge surrounding has the octahedral symmetry owing to the
NaCl-type of structure of NiO. Our Hamiltonian for NiO consists of
two terms: the single-ion-like term $H_{d}$ of the 3$d^{8}$ system
and the d-d intersite spin-dependent term. Calculations somehow
resemble those performed for rare-earth systems, see e.g. Ref.
\cite{7} and they have been recently applied successfully to 3$d$
compounds \cite{8,9}. For the calculations of the quasi-atomic
single-ion-like Hamiltonian of the 3$d^{8}$ system we take into
account the crystal-field interactions of the octahedral symmetry
and the spin-orbit coupling (the octahedral CEF parameter
$B_{4}$=+2 meV - its sign is directly
related to the oxygen anion surroundings, the spin-orbit coupling constant $%
\lambda _{s-o}$ = -41 meV \cite{6}, p. 399). The single-ion states
under the octahedral crystal field and the spin-orbit coupling
have been calculated by consideration of the Hamiltonian:

\begin{figure}[t]
\begin{center}
\includegraphics[width = 7.5 cm]{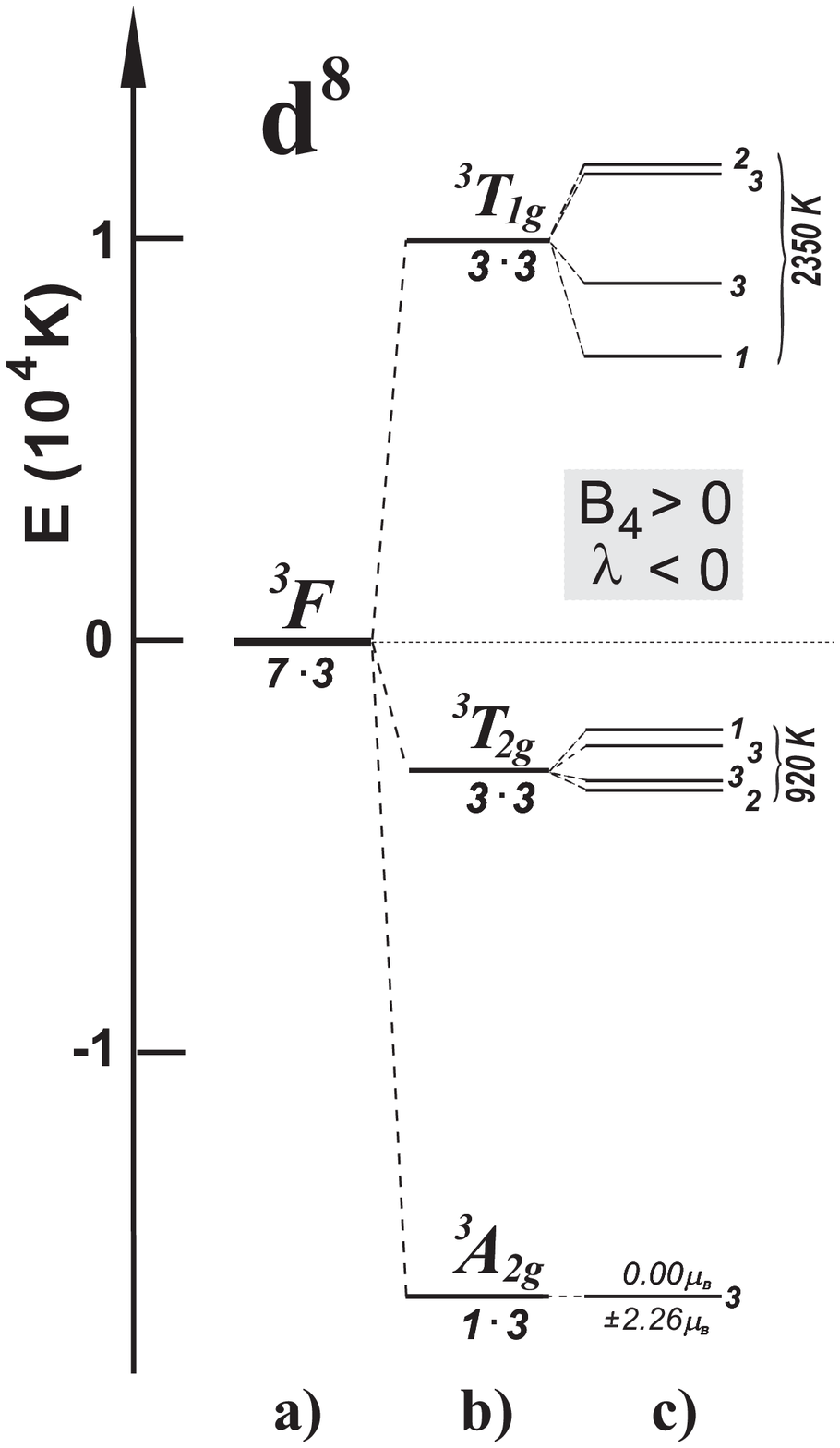}
\end{center} \vspace {-0.9cm}
\caption{The calculated fine electronic structure of the highly-correlated 3$%
d^{8}$ electronic system. a) the 21-fold degenerated $^{3}F$ term
given by two Hund's rules: $S$=1 and $L$=3. b) the effect of the
cubic octahedral crystal-field, c) the combined action of the
spin-orbit coupling and the octahedral crystal field: $B_{4}$=+2
meV, $\lambda _{s-o}$= -41 meV.}
\end{figure}
\vspace {-0.9cm}
\begin{equation}
H_{d}=B_{4}(O_{4}^{0}+5O_{4}^{4})+\lambda _{s-o}L\cdot S
\end{equation}

These calculations have revealed the existence of the fine
electronic structure, Fig. 1, with the charge-formed states
containing three groups of localized states. The higher groups are
at 1.2 and 2.1 eV. For low- and room-temperature properties the
lowest triplet, originating from the cubic subterm $^{3}A_{2g}$,
is the most important as the higher states are not thermally
populated. The triplet states are characterized by the total
moment of 0 and $\pm $2.26 $\mu _{B}$. For the doublet the orbital
moment amounts to $\pm $0.27 $\mu _{B}$. It, however, fully
cancels in the paramagnetic state and reveals itself only in the
presence of the magnetic field, external or internal in the case
of the magnetically-ordered state. The magnetic field polarizes
two states of the doublet. The intersite spin-dependent
interactions cause the (antiferro-)magnetic ordering. They have
been considered in the mean-field approximation with the
molecular-field coefficient {\it n} acting between magnetic
moments $m_{d}$=($L$+2$S$) $\mu _{B}$. The value of $n$ in the
Hamiltonian \vspace{-0.3cm}
\begin{equation} H_{d-d}=n\left( -m_{d}\cdot
m_{d}+\frac{1}{2}\left\langle m_{d}^{2}\right\rangle \right)
\end{equation}
has been adjusted in order to reproduce the
experimentally-observed Neel temperature. The fitted value of $n$
has been found to be -12.2 meV/ $\mu _{B}^{2}$ (=210 T/$\mu
_{B}$). It means that the Ni ion in the magnetically-ordered state
experiences the molecular field of 518 T (at 0 K).
\begin{figure}[t]
\begin{center}
\includegraphics[width = 7.1 cm]{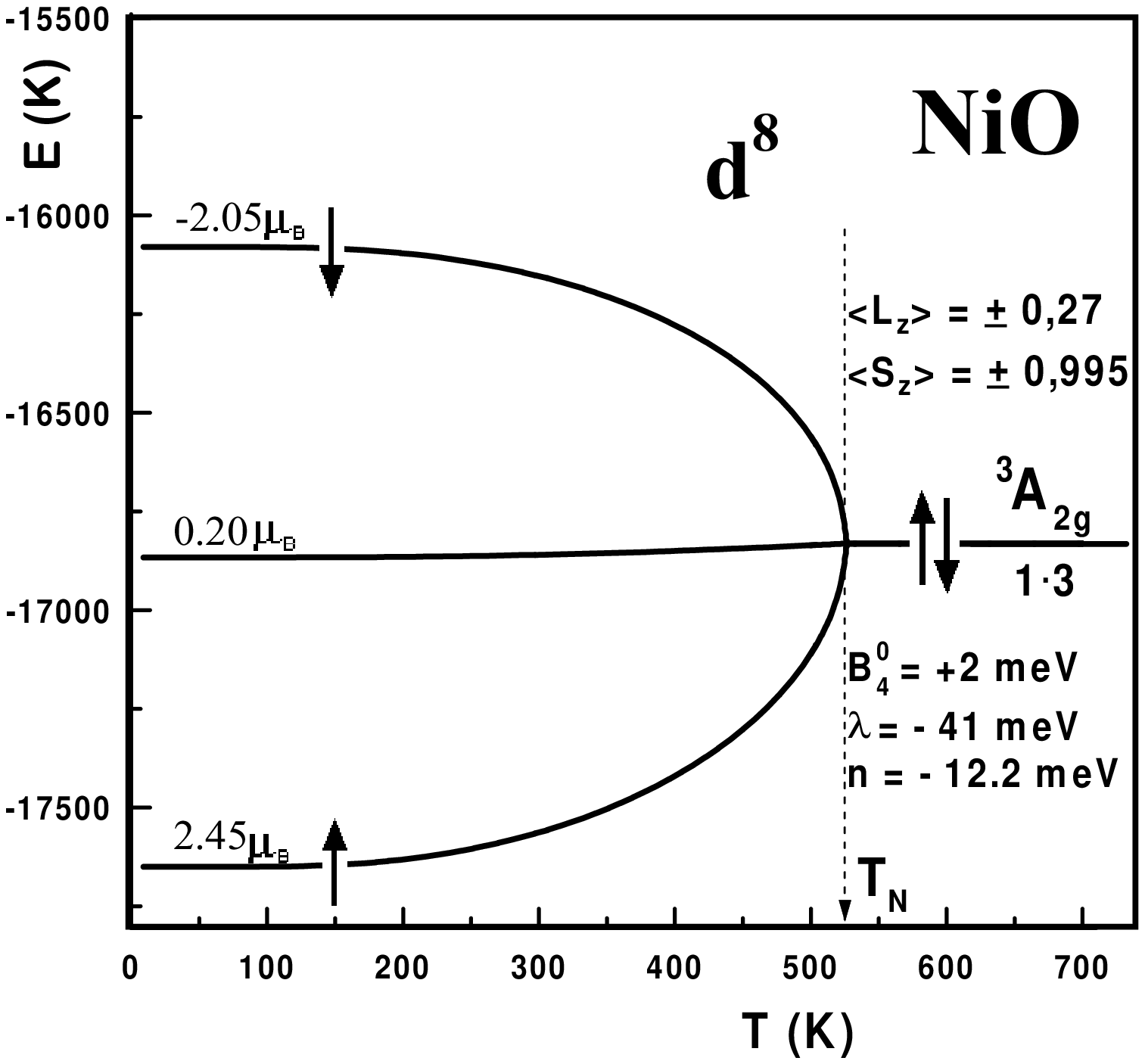}
\end{center}\vspace {-0.9cm}
\caption{The calculated temperature dependence of the energy of
the three lowest levels showing the splitting of the triplet in
the magnetic state (below $T_{N}$=525 K).}
\end{figure}

{\bf 3. Results and discussion}

The calculated value of the magnetic moment at 0 K in the
magnetically-ordered state amounts to 2.45 $\mu _{B}$, Fig. 2. It
is built up from the spin moment $m_{s}$ of 1.99 $\mu _{B}$
(S$_{z}$=0.995) and the orbital moment $m_{o}$ of 0.46 $\mu _{B}$,
Fig. 3. The increase of $m_{o}$ in comparison to the paramagnetic
state, $\pm $ 0.27 $\mu _{B}$, is caused by the further
polarization of the ground-state eigenfunction by the internal
molecular magnetic field. The orbital moment is quite substantial
being about 20\% of the total moment. Our theoretical outcome,
revealing the substantial orbital moment is in nice agreement with
the very recent
experimental result of 2.2$\pm $0.3 $\mu _{B}$ for the Ni moment at 300 K %
\cite{10,11}. This magnetic x-ray experiment has revealed the orbital moment
of 0.32$\pm $0.05 $\mu _{B}$ and the spin moment of 1.90$\pm $0.20 $\mu _{B}$
at 300 K. From the calculated temperature dependence of the total moment,
shown in Fig. 3, one sees that the calculated by us moment at 300 K amounts
to 2.20 $\mu _{B}$ ($m_{s}$= 1.79 $\mu _{B}$, $m_{o}$= 0.41 $\mu _{B}$ )
fully reproducing the experimental result. The ground state triplet is
further split by the trigonal distortion and/or by the internal molecular
field as shown in Fig. 2. Effect of the small trigonal distortion
experimentally observed will be discussed elsewhere. It turns out that the
trigonal distortion is important for the detailed formation of the AF
structure and the direction of the magnetic moment but it only slightly
influences the spin and orbital moments.

\begin{figure}[t]
\begin{center}
\includegraphics[width = 7.0 cm]{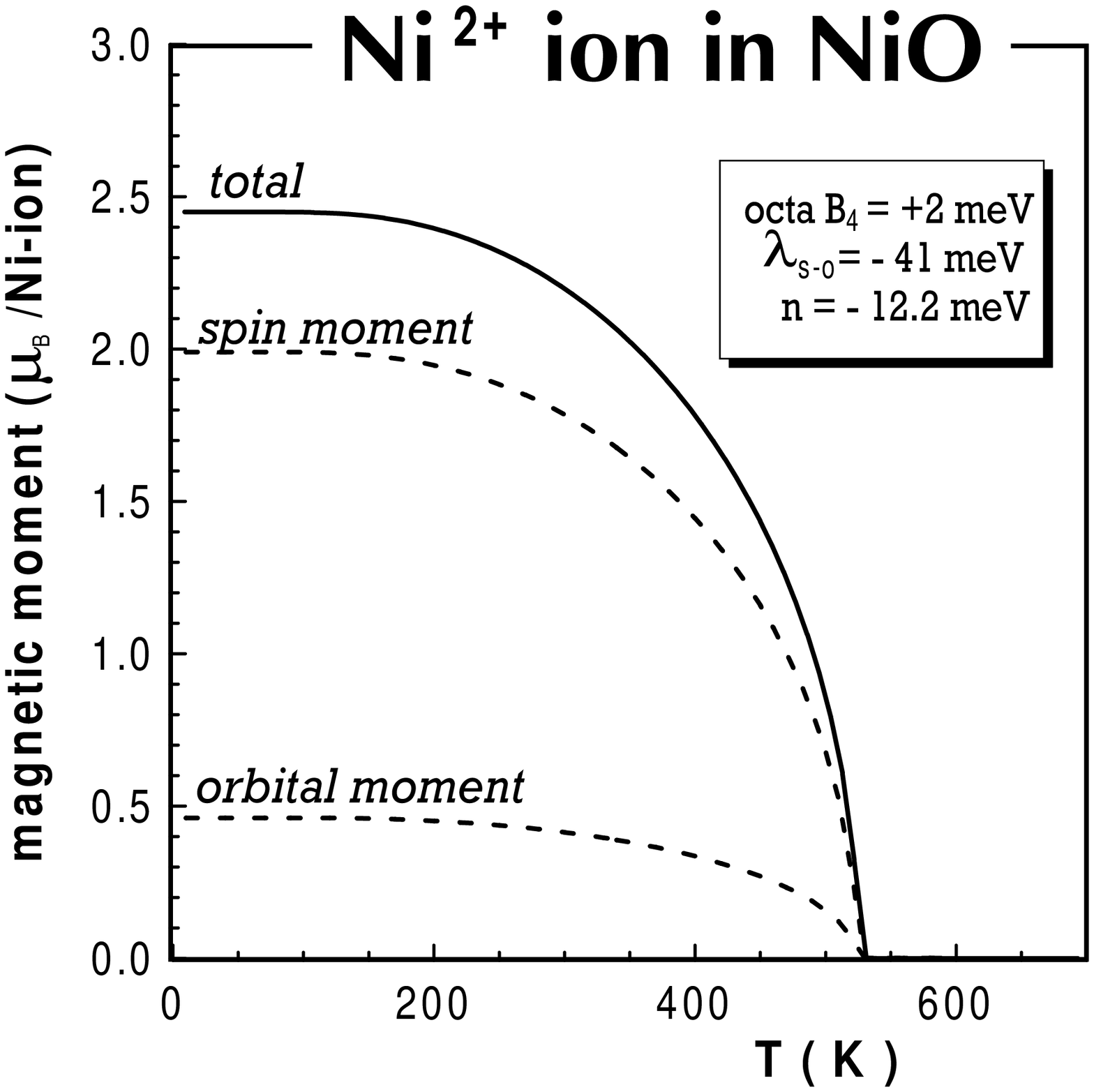}
\end{center}\vspace {-0.9cm}
\caption{The calculated temperature dependence of the
Ni$^{2+}$-ion moment in NiO. At 0 K the total moment of 2.45 $\mu
_{B}$ is built up from the orbital and spin moment of 0.46 and
1.99 $\mu _{B}$. The calculations have been
performed for the quasi-atomic parameters of the octahedral crystal field $%
B_{4}$= +2 meV, the spin-orbit coupling constant $\lambda _{s-o}$=
-41 meV and intersite spin-dependent interactions given by the
molecular-field coefficient $n$ = -12.2 meV/ $\mu _{B}^{2}$.}
\end{figure}

We would like to point out that the evaluation of the orbital
moment is possible provided the intra-atomic spin-orbit coupling
is taken into account. It confirms the importance of the
spin-orbit coupling for the description of the 3d-ion compounds
despite of relative weakness of the spin-orbit coupling. The
present model allows, apart of the ordered moment and its spin and
orbital components, to calculate many physically important
properties like temperature dependence of the magnetic
susceptibility, temperature dependence of the heat capacity (shown
in Fig. 4), the spectroscopic $g$ factor, the fine electronic
structure in the energy window below 3 eV with at least 20
localized states, Fig. 1. The spike-like peak at $T_{N}$ is in
perfect agreement with experimental data \cite{12} obtained on a
single-crystal specimen that yields ''very large, very narrow peak
of 65 cal/Kmol'' \cite{13}. Moreover, we have got that the
magnetically-ordered state of NiO has lower energy than the
paramagnetic one by 3.4 kJ/mol (= 35 meV/ion) at 0 K. Of course,
the energies of magnetic and paramagnetic states become equal at
$T_{N}$.
\begin{figure}[t]
\begin{center}
\includegraphics[width = 6.3 cm]{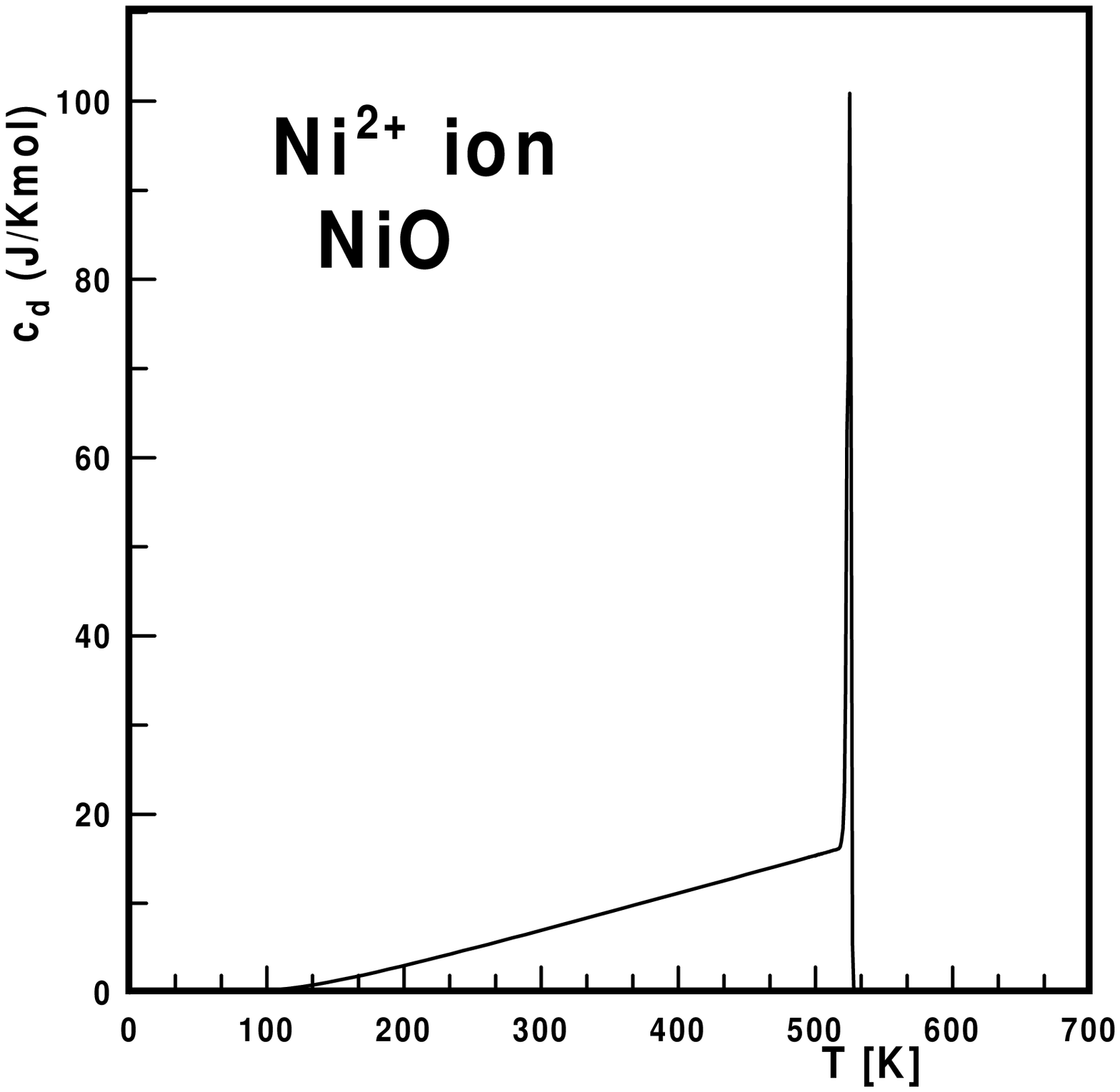}
\end{center}\vspace {-0.9cm}
\caption{The calculated temperature dependence of the 3$d$
contribution to
the heat capacity of NiO. The spike peaks to 214 J/Kmol. This calculated $%
c_{d}(T)$ dependence is in nice agreement with experimental data
shown of Refs 12 and 13, where the spike reaches a value of 272
J/Kmol.}
\end{figure}

Finally, we would like to point out that our approach should not be
considered as the treatment of an isolated ion only - we consider the Ni$%
^{2+}$ ion in the oxygen octahedron. The physical relevance of our
calculations to macroscopic NiO is obvious - the NaCl structure is built up
from the edge sharing Ni$^{2+}$ octahedra.

{\bf 4. Conclusions}

The orbital and spin moment of the Ni$^{2+}$ ion in NiO has been
calculated within the quasi-atomic approach. The orbital moment of
0.46 $\mu _{B}$ amounts at 0 K in the magnetically-ordered state,
to about 20\% of the total moment (2.45 $\mu _{B}$). For this
theoretical outcome, being in nice agreement with the recent
experimental finding, taking into account the intra-atomic
spin-orbit coupling is indispensable. In our atomic approach we
take two Hund's rules to be valid for description of the ground
state of the Ni$^{2+}$ ion - it means that according to our model
the Ni atoms preserve their atomic electronic structure also being
the full part of a solid. The presented approach explains in a
very natural way the insulating state of NiO - in fact, NiO is one
of the best insulators. Good description of many physical
properties indicates that 3$d$ electrons in NiO are in the
extremely strongly-correlated limit. In contrary to many
considerations yielding a continuous electronic structure our
atomic approach yields the discrete energy states for
3$d$-electrons in NiO and is in the thinking line of the atomic
considerations of Refs. \cite{10} and \cite{11}. As an extra
result, these studies have revealed that the Ni$^{2+}$ ion is the
Jahn-Teller ion.

A note added 17 December 2004. A very short version of this paper
has been published in Acta Phys. Pol. A {\bf 97} (2000) 963, only
with Fig. 3 \cite {14,15}. Readers interested in this subject are
encouraged to contact us and our papers on "Orbital moment in CoO"
\cite{16} and "$^{5}D$ term origin of the excited triplet in
LaCoO$_{3}$" \cite{17}. Our "Strongly-correlated crystal-field
approach to 3d oxides - the orbital magnetism in 3d-ion compounds" one can found as Ref. 18. \\
$^\spadesuit$This paper has been submitted 11 July, 2001 to Phys.
Rev. B and is registered under the number BG8156 being protected
for the author rights.

\end{document}